\documentclass[runningheads,a4paper]{llncs}

\usepackage{graphicx}
\usepackage{footnote}

\usepackage{url}
\urldef{\mailsa}\path|{name.surname}@barcelonamedia.org|
\newcommand{\keywords}[1]{\par\addvspace\baselineskip
\noindent\keywordname\enspace\ignorespaces#1}

\begin{document}

\mainmatter 
\title{Not all paths lead to Rome:\\Analysing the network of sister cities}

\titlerunning{Not all paths lead to Rome: Analysing the network of
  sister cities}

\author{Andreas Kaltenbrunner%
  \thanks{This work was supported by the Spanish CDTI under the CENIT
    program, project CEN-20101037, and ACC1\'O -Generalitat de
    Catalunya (TECRD12-1-0003).}%
  \and Pablo Arag\'on\and David Laniado\and Yana Volkovich%
  \thanks{Acknowledges the Torres Quevedo Program from the Spanish
    Ministry of Science and Innovation, co-funded by the European
    Social Fund.}%
}
\authorrunning{A.~Kaltenbrunner \and P.~Arag\'on \and D.~Laniado \and
  Y.~Volkovich
}
\institute{Barcelona Media Foundation,
Barcelona, Spain\\
\mailsa}

\toctitle{Not all paths lead to Rome: Analysing the network of sister
  cities}
\tocauthor{Kaltenbrunner, Aragon,
  Laniado and Volkovich}
\maketitle

\begin{abstract}
  This work analyses the practice of sister city pairing. We
  investigate structural properties of the resulting city and country
  networks and present rankings of the most central nodes in these
  networks. We identify different country clusters and find that the
  practice of sister city pairing is not influenced by geographical
  proximity but results in highly assortative networks.
  \keywords{Social network analysis, sister cities, social
    self-organisation}
\end{abstract}

\section{Introduction}\label{sec:intro}
Human social activity in the form of person-to-person interactions has
been studied and analysed in many contexts, both for
online~\cite{Mislove2007} and off-line
behaviour~\cite{wasserman1994social}. However, sometimes social
interactions give rise to relations not anymore between individuals
but rather between entities like companies~\cite{hopner2004politics},
associations~\cite{moore2003international} or even
countries~\cite{caldarelli2012}.  Often these relations are associated
with economic exchanges~\cite{caldarelli2012}, sports
rivalry~\cite{mukherjee2012identifying} or even
cooperation~\cite{moore2003international}.

In this work we study one type of such relations expressed in the form
of \emph{sister city partnerships}\footnote{Sometimes the same concept
  is also referred to as \emph{twin town}, \emph{partnership town},
  \emph{partner town} or \emph{friendship town}. Here we use
  preferentially the term \emph{sister city}.}.  The concept of sister
cities refers to a partnership between two cities or towns with the
aim of cultural and economical exchange. Sometimes these partnerships
are also generated as a platform to support democratic processes.
Most partnerships connect cities in different countries, however also
intra-country city partnerships exist.

We extracted the network of sister cites as reported on the English
Wikipedia, as far as we know the most extensive but certainly not
complete collection of this kind of relationships. The resulting social
network, an example of social self organisation, is analysed in its
initial form and aggregated per country. Although there exist studies
that analyse networks of cities (e.g. networks generated via
aggregating individual phone call
interactions~\cite{krings2009scaling}) to the best of our knowledge
this is the first time that institutional relations between cities are
analysed.

\section{Dataset description}
\label{sec:methods}
The dataset used in this study was constructed (using an automated
parser and a manual cleaning process) from the listings of sister
cities on the English Wikipedia.\footnote{Starting from {\scriptsize
    \url{http://en.wikipedia.org/wiki/List_of_twin_towns_and_sister_cities}},
  which includes links to listings of sister cities grouped by
  continent, country and/or state.}  We found 15~225 pairs of sister
cities, which form an undirected\footnote{Although only 29.8\% of the
  links were actually reported for both directions.}  \emph{city
  network} of 11~618 nodes. Using the \verb|Google Maps API| we were able
to geo-locate 11~483 of these cities.

We furthermore construct an aggregated undirected and weighted
\emph{country network}, where two countries $A$ and $B$ are connected
if a city of country $A$ is twinned with a city of country $B$. The
number of these international connections is the edge weight.  The
country network consists of 207 countries and 2~933 links. Some
countries have self-connections (i.e. city partnerships within the
same country).  Germany has the largest number of such self links as a
result of many sister city relations between
the formerly separated East and West Germany.

\begin{table}[!tb]
  \caption{ Network properties: number of nodes $N$ and edges $K$,
    average clustering coefficient $\langle C \rangle$, \% of nodes in
    the giant component GC, average path-length $\langle d \rangle$.}
      \label{tab:NW_measures}
      \centering
      \small
      \begin{tabular}{@{}l|rrrrcc@{}}\hline
        network & $N$~~~ & $K$~~~ & $\langle C \rangle$~ &\% GC~ &  $\langle d \rangle$\\
        \hline
        \hline
        city network & $11\;618$~ & $15\;225$~ & $0.11$~ & $61.35$\%~ & $6.74$ \\
        country network & $207$~ & $2933$~ & $0.43$~ & $100$\%~ & $2.12$ \\
        \hline
      \end{tabular}
\end{table}

Table~\ref{tab:NW_measures} lists the principal social network
measures of these two networks. The clustering coefficient of the city
network is comparable to the values observed in typical social
networks~\cite{newman2002random}.  Also the average path-length
between two nodes nodes is with $6.7$ in line with the famous
six-degrees-of-separation.  The country network is denser, witnessed
by the remarkably high value of the clustering coefficient ($\langle C
\rangle=0.43$), and a very short average distance of $2.12$.

\begin{figure}[!tb]
  \centering
  \includegraphics[width=.4\textwidth]{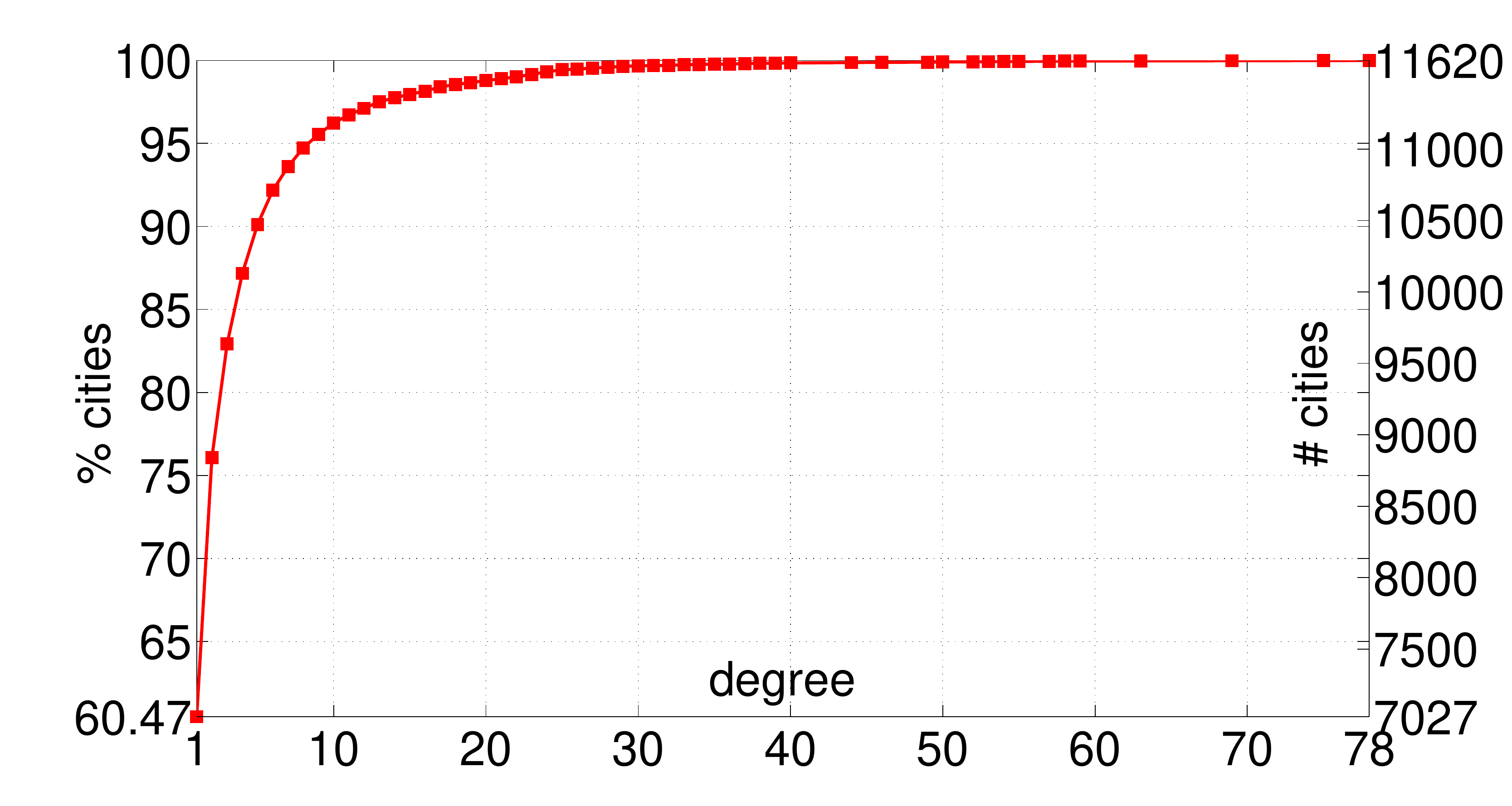}
  \includegraphics[width=.46\textwidth]{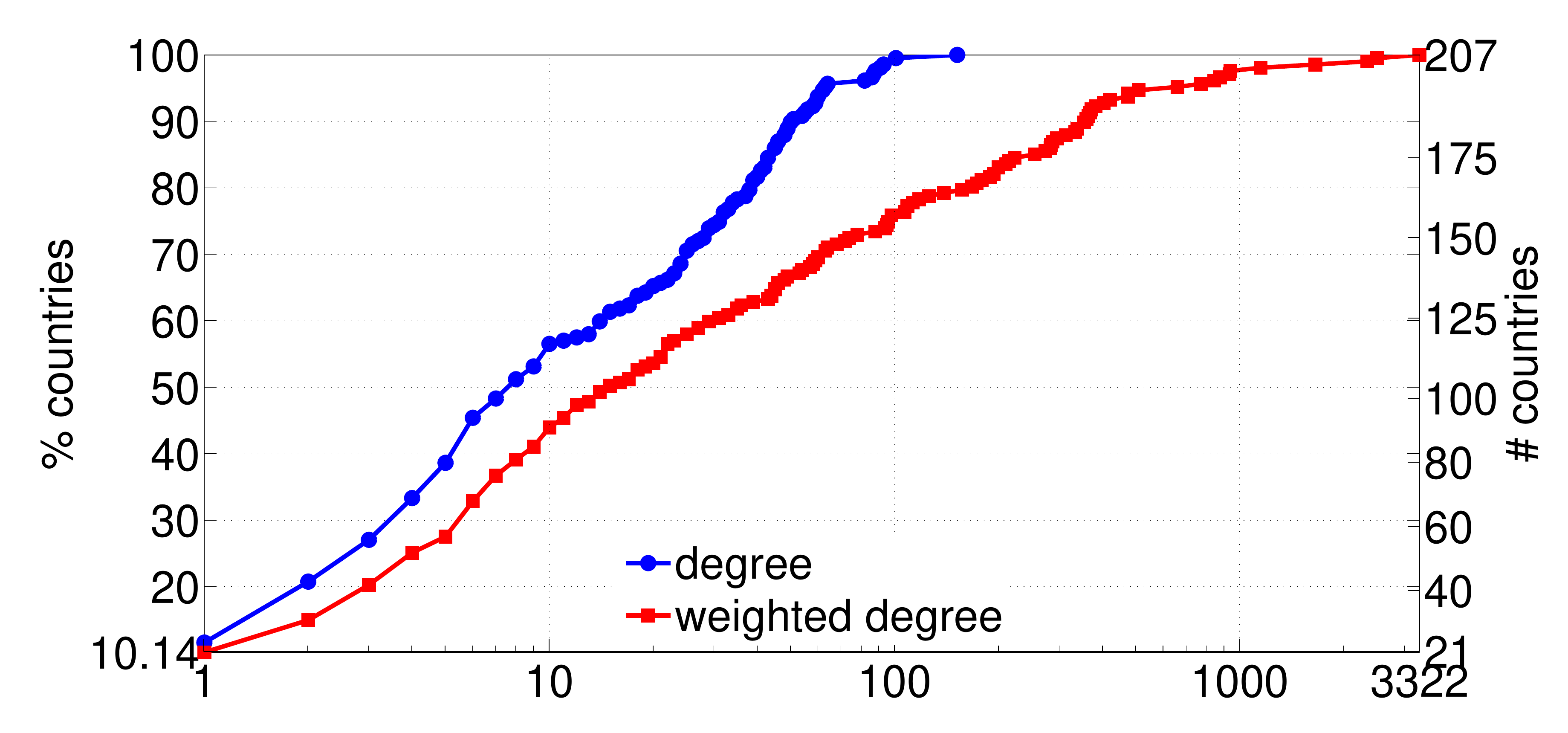}
\vspace{-4mm}
  \caption{Cumulative degree distribution in the city (left) and country
    networks (right). \label{fig:degree}}
\end{figure}

In Figure~\ref{fig:degree} we plot the degree distributions of both
networks. We observe in Figure~\ref{fig:degree} (left) that more than
60\% of the cities have only one sister city, about 16\% have two and
only less than 4\% have more than 10.  For the countries we observe in
Figure~\ref{fig:degree} (right) that around 58\% of the countries have
less than 10 links to other countries, but at the same time more than
20\% of the countries have more than 100 sister city connections
(i.e. weighted degree $\geq100$).  Both networks have skewed
degree-distributions with a relative small number of hubs.

\begin{savenotes}
\begin{table}[!tb]
  \caption{Comparing assortativity coefficients $r$ of the city
    network with the mean assortativity coefficients $r_{rand}$ and
    the corresponding stdv $\sigma_{rand}$ of randomised
    networks. Resulting Z-scores $\geq 2$ (in bold) indicate
    assortative mixing. Apart from the city degrees,
    the city properties used coincide with the corresponding country indexes.} 
   \label{tab:und_asso}
   \centering
   \small
   \begin{tabular}{@{}l|rrrr@{}}
     \hline
     property &$r$&$r_{rand}$&$\sigma_{rand}$&$Z$\\\hline
     \hline
     city degree & 0.3407&-0.0037&0.0076&\bf{45.52}\\
     Gross Domestic Product (GDP)\footnote{Source \scriptsize \url{http://en.wikipedia.org/wiki/List_of_countries_by_GDP_(nominal)}}&0.0126 & -0.0005 & 0.0087 & 1.51\\
     GDP per capita\footnote{Source: \scriptsize \url{http://en.wikipedia.org/wiki/List_of_countries_by_GDP_(nominal)_per_capita}}&0.0777 & 0.0005 & 0.0078 & \bf{9.86}\\
     Human Development Index (HDI)\footnote{Source: \scriptsize \url{http://en.wikipedia.org/wiki/List_of_countries_by_Human_Development_Index}}&0.0630 & -0.0004 & 0.0075 & \bf{8.46}\\
     Political Stability Index\footnote{Source: \scriptsize \url{http://viewswire.eiu.com/site_info.asp?info_name=social_unrest_table}}&0.0626 & 0.0004 & 0.0090 & \bf{6.94}\\
     \hline
   \end{tabular}
\end{table}
\end{savenotes}

\section{Assortativity}\label{sec:ass}
To understand mixing preferences between cities, we follow the
methodology of~\cite{foster2010edge} and calculate an assortativity
measure based on the Z-score of a comparison between the original
sister city network and 100 randomised equivalents.  For
degree-assortativity, randomised networks are constructed by
reshuffling the connections and preserving the degree; in the other
cases, the network structure is preserved while the values of node
properties are reshuffled.

Table~\ref{tab:und_asso} gives an overview of the results. We find
that the city network is highly assortative indicating a clear
preference for connections between cities with similar degree.  We
also analyse assortativity scores for other variables and find that
cities from countries with similar Gross Domestic Product (GDP) per
capita, Human Development Index or even similar indexes of political
stability are more likely to twin. Only for the nominal GDP neutral
mixing is observed.

\section{Rankings}\label{sec:rankings}
We discuss now city and country rankings based on centrality measures.
For the sister city network we show the top 20 cities ranked by degree
(Table~\ref{tab:top_degree}, left).  Saint Petersburg, often referred
to as the geographic and cultural border of the West and East, is the
most connected and also most central sister city. There are also
cities, such as Buenos Aires, Beijing, Rio de Janeiro and Madrid,
which have large degrees but exhibit lower betweenness ranks. In
particular, the Spanish and the Chinese capitals have significantly
lower values of betweenness, which could be caused by the fact that
other important cities in these countries (e.g. Barcelona or Shanghai)
act as primary international connectors.

In Table~\ref{tab:top_degree} (right) we present rankings for the
country network. In this case the USA lead the rankings in the two
centrality measures we report. The top ranks are nearly exclusively
occupied by Group of Eight (G8) countries suggesting a relation
between economic power and sister city connections.

\begin{table}[!tb]
  \caption{The top 20 cities (left) and countries (right) ranked by
    (weighted) degree. Ranks for betweenness centrality in
    parenthesis.} \label{tab:top_degree}
  \centering
  \scriptsize{
\begin{tabular}{@{}l|r|r@{\hspace{4pt}}rc@{}}
\hline
    city & degree & \multicolumn{2}{|c}{betweenness}  \\
    \hline \hline
Saint Petersburg & 78 & 1 562 697.97 & (1)\\
Shanghai & 75 & 825 512.69 & (4)\\
Istanbul & 69 & 601 099.50 & (12)\\
Kiev & 63 & 758 725.12 & (5)\\
Caracas & 59 & 430 330.45 & (23)\\
Buenos Aires & 58 & 348 594.25 & (36)\\
Beijing & 57 & 184 090.42 & (124)\\
S\~ao Paulo & 55 & 427 457.92 & (24)\\
Suzhou & 54 & 740 377.17 & (6)\\
Taipei & 53 & 486 042.21 & (20)\\
Izmir & 52 & 885 338.70 & (3)\\
Bethlehem & 50 & 1 009 707.96 & (2)\\
Moscow & 49 & 553 678.88 & (16)\\
Odessa & 46 & 724 833.39 & (8)\\
Malchow & 46 & 519 872.56 & (17)\\
Guadalajara & 44 & 678 060.06 & (9)\\
Vilnius & 44 & 589 031.92 & (14)\\
Rio de Janeiro & 44 & 381 637.67 & (29)\\
Madrid & 40 & 135 935.80 & (203)\\
Barcelona & 39 & 266 957.88 & (60)\\
\hline
\end{tabular}
\quad
\begin{tabular}{@{}l|r|r@{\hspace{4pt}}r@{}}
\hline
    country & weighted degree & \multicolumn{2}{|c}{betweenness} \\
    \hline \hline
USA & 4520 & 9855.74 & (1)\\
France & 3313 & 1946.26 & (3)\\
Germany & 2778 & 886.78 & (6)\\
UK & 2318 & 2268.32 & (2)\\
Russia & 1487 & 483.65 & (9)\\
Poland & 1144 & 34.09 & (33)\\
Japan & 1131 & 168.47 & (20)\\
Italy & 1126 & 849.20 & (7)\\
China & 1076 & 1538.42 & (4)\\
Ukraine & 946 & 89.22 & (27)\\
Sweden & 684 & 324.84 & (14)\\
Norway & 608 & 147.06 & (22)\\
Spain & 587 & 429.79 & (11)\\
Finland & 584 & 30.24 & (35)\\
Brazil & 523 & 332.26 & (13)\\
Mexico & 492 & 149.70 & (21)\\
Canada & 476 & 72.01 & (28)\\
Romania & 472 & 34.44 & (32)\\
Belgium & 464 & 145.18 & (23)\\
The Netherlands & 461 & 274.79 & (16)\\
\hline
\end{tabular}
}
\end{table}

\section{Clustering of the country network}\label{sec:visualisations}
\begin{figure}[!tb]
  \centering
  \includegraphics[angle=90,width=\textwidth]{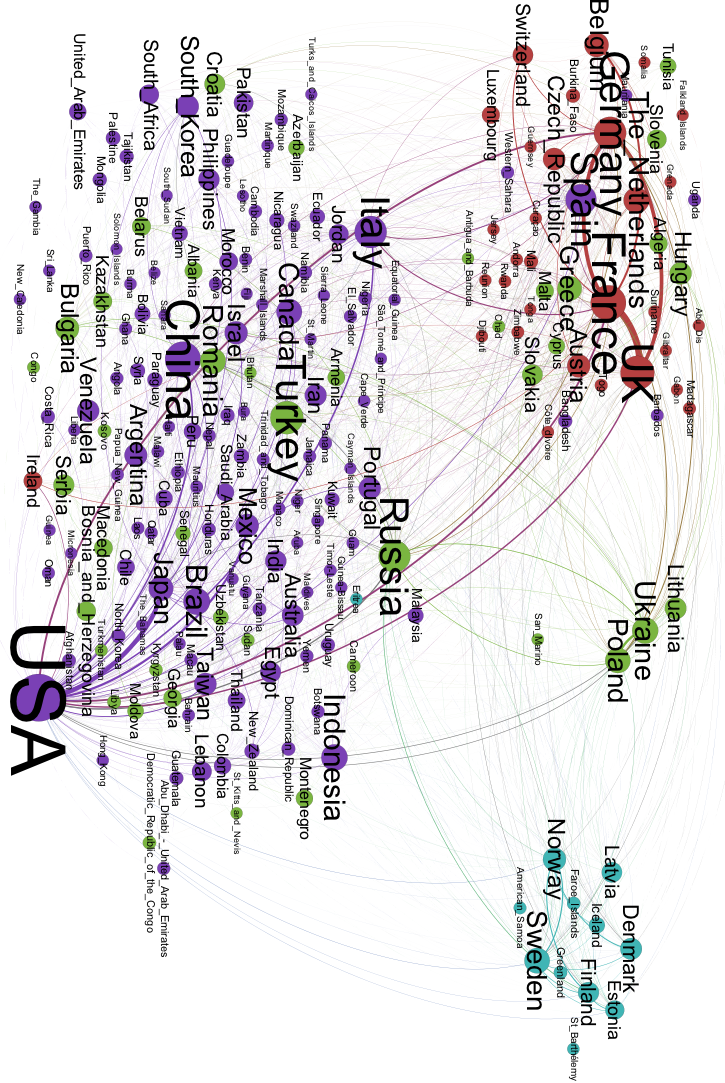}
  \caption{Country network: node size corresponds to degree and node
    colours indicate the four clusters obtained with the Louvain
    method. }  \label{fig:country_network_global}
\end{figure}

In Figure~\ref{fig:country_network_global} we depict the country
network. Node size corresponds to the weighted degree, and the width
of a connection to the number of city partnerships between the
connected countries. The figure shows the central position of
countries like the USA, France, UK and China in this network.

The colours of the nodes correspond to the outcome of node clustering
with the Louvain method. We find 4 clusters. The largest one (in
violet) includes the USA, Spain and most South American, Asian, and
African countries. The second largest (in green) is composed of
Eastern-European and Balkan countries: Turkey, Russia, and Poland are
the most linked among them. The third cluster (in red) consists of
Central and Western-European countries and some of their former
colonies. It is dominated by Germany, UK, France and the
Netherlands. Finally, the smallest cluster (in cyan) mainly consists
of Nordic countries.

The clustering suggests cultural or geographical proximity being 
a factor in city partnerships. In the next section we will investigate this
further.

\section{Distances}\label{sec:distances}
To test the extent to which geographical proximity is a important
factor for city partnership we analyse the distribution of
geographical distances between all pairs of sister cities.

\begin{figure}[!tb]
  \centering
  \includegraphics[width=.49\columnwidth]{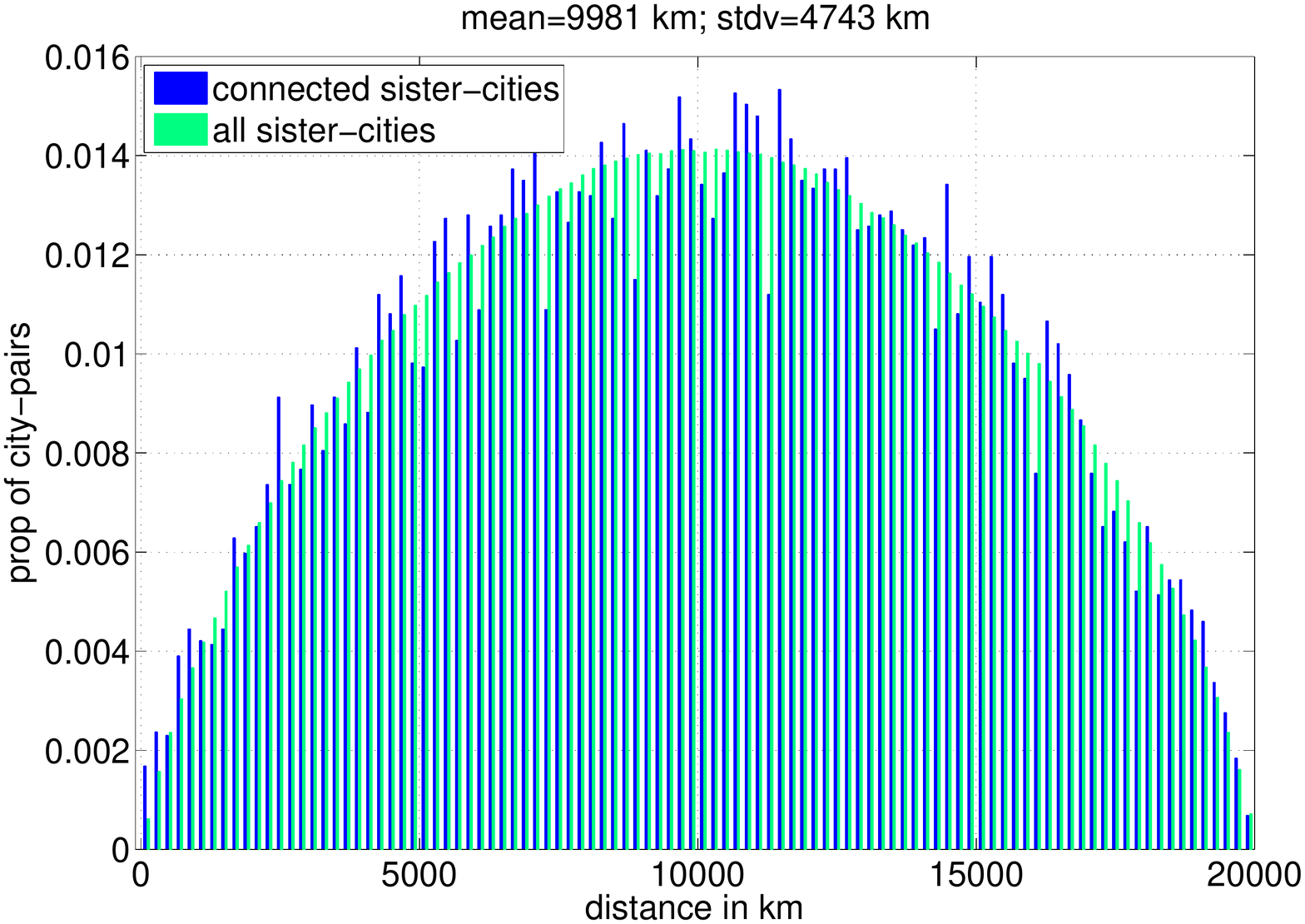}
  \includegraphics[width=.49\columnwidth]{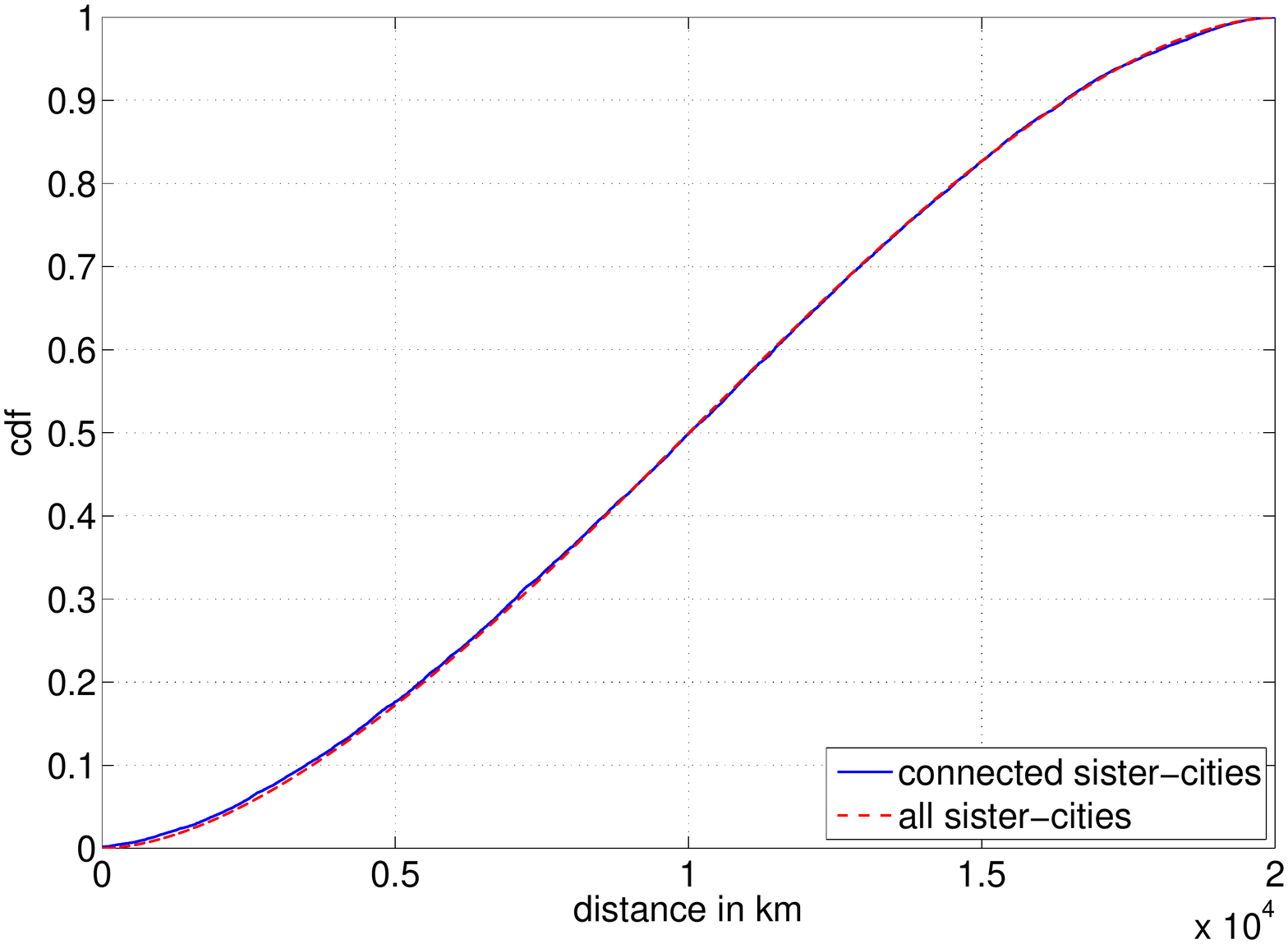}
  \caption{Distribution of distances between sister cities. The
    distribution is practically identical to overall distance distribution
    between the cities.}
  \label{fig:city_distances_distribution}
\end{figure}

\begin{figure}[!tb]
    \centering
   \includegraphics[angle=90,width=.95\textwidth]{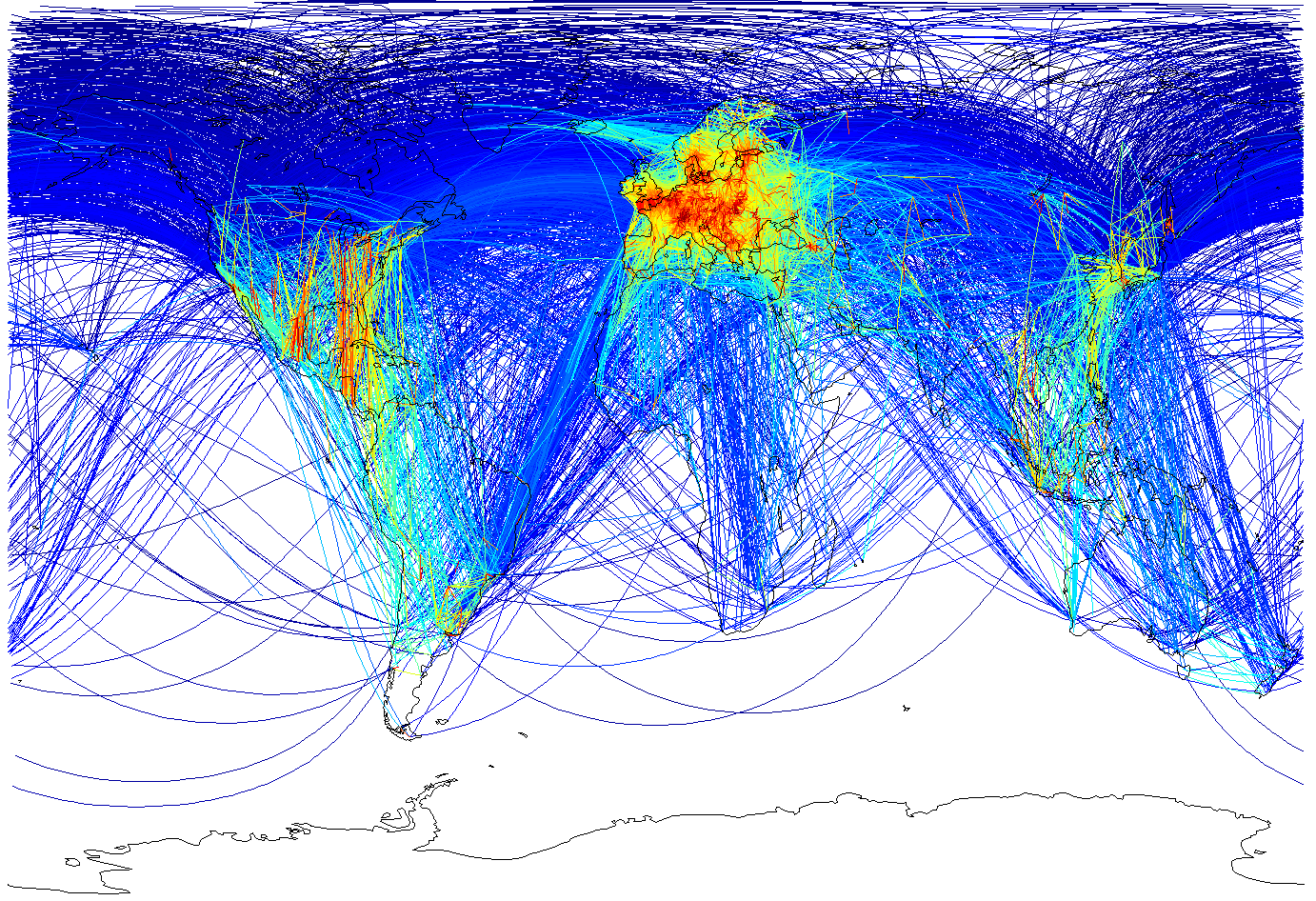}
   \caption{Connections between sister cities visualised on a world
      map. Shorter connections are drown in a brighter colour.}
    \label{fig:city_network_global}
  \end{figure}

Figure~\ref{fig:city_distances_distribution} depicts this distribution
as a histogram (blue bars in the left sub-figures) or as a cumulative
distribution (blue line in the right sub-figure). The figure also
shows the overall distance distribution between all geo-located sister
cities (green bars and red line). We observe that there is nearly no
difference (apart from some random fluctuations) between these two
distributions. The fluctuations get cancelled out in the cumulative
distributions where the two curves are nearly overlapping. Only for
very short distances the likelihood of city partnership with close
sites is a bit larger than random. This can also be observed in the
very small difference between the average distance of two randomly
chosen sister cities ($10~006$~km) and two connected sister cities
($9~981$~km). Figure~\ref{fig:city_network_global} visualises the
distances between sister cities overlaid over a World map.

\section{Conclusions}\label{sec:conclusion}
We have analysed the practice of establishing sister city connections
from a network analysis point of view.  Although there is no guarantee
that our study covers all existing sister city relations, we are
confident that the results obtained give reliable insights into the
emerging network structures and country preferences

We have found that sister city relationships reflect certain
predilections in and between different cultural clusters, and lead to
degree-assortative network structures comparable to other types of
small-world social networks. We also observe assortative mixing with
respect to economic or political country indexes.

The most noteworthy result may be that the geographical distance has
only a negligible influence when a city selects a sister city.  This is
different from what is observed for person-to-person social
relationships (see for example~\cite{kaltenbrunner2012WOSN}) where the
probability of a social connection decays with the geographical
distance between the peers. It may, thus, represent the first
evidence in real-world social relationships (albeit in its
institutional form) for the death of distance, predicted originally as
a consequence of decrease of the price of human
communication~\cite{cairncross2001death}.

Possible directions for future work include combination of the
analysed networks with the networks of air traffic or goods exchange
between countries.

\end{document}